\documentclass{article}

\usepackage{PRIMEarxiv}

\usepackage[utf8]{inputenc} 
\usepackage[T1]{fontenc}    
\usepackage{hyperref}       
\usepackage{url}            
\usepackage{booktabs}       
\usepackage{amsfonts}       
\usepackage{nicefrac}       
\usepackage{microtype}      
\usepackage{lipsum}
\usepackage{fancyhdr}       
\usepackage{graphicx}       
\graphicspath{{media/}}     
\usepackage{xcolor}
\usepackage{pifont}
\newcommand{\cmark}{\textcolor{green}{\ding{51}}}%
\newcommand{\xmark}{\textcolor{red}{\ding{55}}}%

\DeclareUnicodeCharacter{0301}{\'e}

\title{MicroLabVR: Interactive 3D Visualization of Simulated Spatiotemporal Microbiome Data in Virtual Reality
}

\author{
  Simon Burbach, Maria Maleshkova \\
  Professorship of Data Engineering \\
  Helmut Schmidt University \\
  Hamburg, Germany \\
  \texttt{\{burbachs, maleshkm\}@hsu-hh.de} \\
  \And
  Florian Centler\thanks{Shared senior authorship} \\
  Bioinformatics, School of Science and Technology \\
  University of Siegen \\
  Siegen, Germany \\
  \texttt{florian.centler@uni-siegen.de} \\
  \And
  Tanja Joan Schmidt\footnotemark[\value{footnote}]\\
  Department of Clinical Psychology\\
  University of Siegen\\
  Siegen, Germany\\
  \texttt{tanja.schmidt@uni-siegen.de} \\
}

\begin{document}
\maketitle

\begin{abstract}
Microbiomes are a vital part of the human body, engaging in tasks like food digestion and immune defense. Their structure and function must be understood in order to promote host health and facilitate swift recovery during disease. Due to the difficulties in experimentally studying these systems \textit{in situ}, more research is being conducted in the field of mathematical modeling. Visualizing spatiotemporal data is challenging, and current tools that simulate microbial communities' spatial and temporal development often only provide limited functionalities, often requiring expert knowledge to generate useful results. To overcome these limitations, we provide a user-friendly tool to interactively explore spatiotemporal simulation data, called MicroLabVR, which transfers spatial data into virtual reality (VR) while following guidelines to enhance user experience (UX). With MicroLabVR, users can import CSV datasets containing population growth, substance concentration development, and metabolic flux distribution data. The implemented visualization methods allow users to evaluate the dataset in a VR environment interactively. MicroLabVR aims to improve data analysis for the user by allowing the exploration of microbiome data in their spatial context.
\end{abstract}

\keywords{Data Visualization \and Microbiome \and Metabolic Modeling \and Virtual Reality}

\section{Introduction}
\label{sec:introduction}
\begin{figure}[h]
    \centering
    \includegraphics[width=0.7\linewidth]{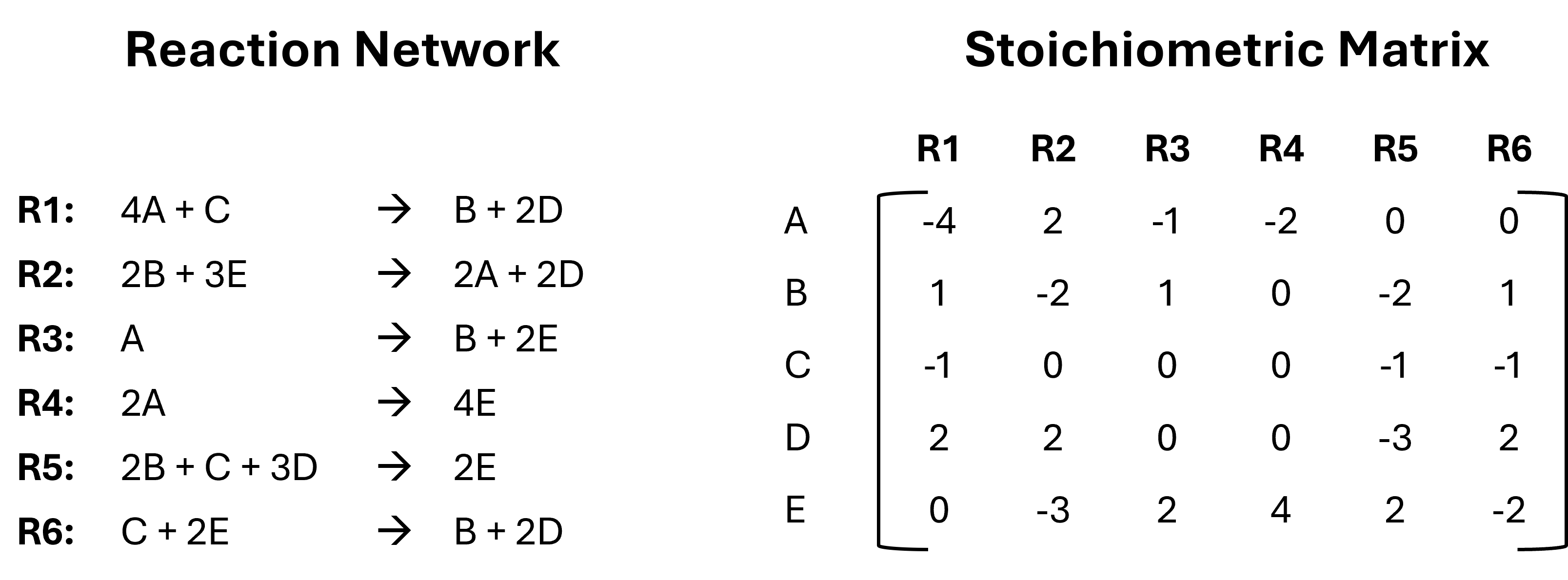}
    \caption[Stoichiometric matrix]{A reaction network with the representing stoichiometric matrix. R$n$ are the reactions, and letters A to E are the metabolites involved in the reactions (modeled after \cite{Singh2013})}
    \label{fig:matrix}
\end{figure}

Life on earth started unicellular, and many higher organisms today, including humans, rely on a symbiosis with microbial communities consisting of diverse unicellular organisms, including bacteria and archaea \cite{Margulis1971, King1977, Sagan1967, Michod2007}. Understanding the complexity and diversity is essential to maintain and restore human and ecological health \cite{Foo2017}.
Current omics methods (metagenomics \cite{Wooley2010, Zhang2021}, functional metagenomics \cite{Mirete2016}, metatranscriptomics \cite{Moran2012, Ojala2023}, metaproteomics \cite{Wang2020}, metabolomics \cite{Kumar2019, Liu2017, Yen2015} etc.) require a lot of laboratory equipment and time to identify the functions and compositions of microbiomes. To make the most of this data, mathematical modeling has emerged as a technique that builds on this data and elucidates mechanisms by enabling experiments that are difficult to perform \textit{in situ} \cite{Kumar2019, Fabien2009, Tomlin2007}. These mathematical models mainly follow two approaches: constraint-based and agent-based. Constraint-based modeling (CBM) focuses on the simulation of metabolic networks, and agent-based modeling (ABM) focuses on the simulation of individuality in cellular populations. In addition, both approaches can be combined \cite{Bauer2017}. CBM  uses a common method for modeling metabolic processes in microbiomes called flux balance analysis (FBA). It can be used to model fluxes of cellular metabolism in order to investigate growth phenotypes and cell-to-cell interactions \cite{Zhuang2010, Stolyar2007, Wintermute2010} and to develop drugs that intervene in the metabolism of pathogens \cite{Lee2010, Raman2005, Lee2009, Lakshmanan2012}. The principle of FBA is based on the representation of biochemical network reactions by a stoichiometric matrix (see Figure \ref{fig:matrix}) detailing the stoichiometry of each reaction with respect to its educt and product metabolites. In addition to the restrictive upper and lower limits for uptake fluxes, which depend on substrate availability in the vicinity of the cellular population, the pursuit of maximum biomass growth is a frequent goal in the models. In these cases, the metabolic pathways of an organism are aligned to optimize growth \cite{Kauffman2003}.

The simulation itself is based on the assumption of a steady state \cite{Lakshmanan2012, Orth2010, Oberhardt2009}. This describes the state of a system, in which the given boundary conditions, the concentrations of metabolites remain constant, i.e., production and uptake balance each other out for each internal metabolite \cite{Kauffman2003}. In steady state models, the kinetic parameters of reactions are therefore neglected. An extension of the model from monoculture to the community level requires calculating an optimization function of the fluxes on several levels (bi-level optimization). In such cases, the fluxes of the individual species are optimized first, followed by the reactions of the entire community \cite{Zomorrodi2012}.\\
In order to represent realistic conditions, dynamics must be considered additionally. To simulate these changes, dynamic FBA (dFBA) models assume that the system consists of several successive \textit{quasi-steady} states. In dFBA, a \textit{quasi-steady} state is calculated for each time step, and the changes between these states are modeled to simulate the system's temporal development. Changes in this context refer to changes in extracellular compounds and changes in biomass concentrations of involved microbial species. Such models enable the analysis of population growth as well as the consumption and production of substances by individual populations \cite{Henson2014}.\\
A final extension considers the spatial development of a microbial community. For this purpose, the space is discretized. It is assumed that there is homogeneous mixing within each grid cell. Diffusion parameters exist between the compartments that enable the exchange of substances among each other \cite{Harcombe2014}. This results in concentration gradients that drive substrate distribution and development in space \cite{Bauer2017, Borer2021}. First, the state in each grid cell is updated, considering the uptake of substances and biomass production. Global processes are then taken into account. Substances can diffuse between the grid cells, and cells can migrate \cite{Harcombe2014}.\\
While FBA models are typically used to solve an optimization problem for populations and often entire communities, agent-based modeling (ABM) is concerned with the simulation of individual organisms in a population.
The focus is on analyzing the behavior and interactions of the participants rather than investigating microbial metabolic processes. With ABM, organisms are considered as individuals, also called agents, and are separate data objects with their own states, to which certain properties, behaviors, and interactions are linked \cite{Bauer2017, Holland1992, Grimm2005, Kaul2013}. This also differs from FBA, which is only based on stoichiometric data. The behavior of agents in the simulation is governed by rules that are assigned to the system. These conditions make it possible to investigate complex dynamics in a population with the properties of the individual organisms. In addition to such dynamics, spatiotemporal processes such as gradients of substrates can also be analyzed \cite{An2009, Shashkova2016}.\\
Current tools for spatial and temporal modeling of microbiomes mainly focus on the ability to calculate microbial development accurately, and less on data visualization. Datasets produced by these modeling tools usually contain multiple dimensions unsuitable to be shown on a two-dimensional (2D) screen \cite{Pirch2021}. Furthermore, not all modeling tools have a graphical user interface (GUI) \cite{Bauer2017, Biggs2013} but require programming knowledge to operate, which limits the usability of the application for many scientists \cite{Wilson2014, Mangul2019, Barone2017}.\\
Virtual Reality (VR) has the potential to interactively visualize multidimensional microbiome datasets in three dimensions by providing GUIs to guide users through the simulation step by step \cite{Donalek2014, Aichem2022}.
VR refers to the simulation and simultaneous perception of an apparent reality and its physical properties, which is generated in real-time on a processing unit. Users wearing a head-mounted display (HMD) can interact with this simulated reality. In VR, all possible scenarios can be depicted, including a more precise representation of a microbiome dataset. During such simulations, the user is, depending on the setting, completely visually and preferably also audibly separated from their real environment to increase the level of immersion \cite{Zheng1998, Jennett2008}.
Serious gaming applications increase engagement and make operating the application more intuitive. This is because serious gaming simulations reflect real-life situations, combined with gamification elements, to additionally increase the learning effect, which refers to the term situated learning\footnote{More about situated learning: \url{https://ijiet.org/papers/48-R017.pdf} (last accessed 10.08.2025)} \cite{Allcoat2018, Mouatt2020, GhavamiHoseinPour2025}.\\
This paper investigates which visualization techniques are particularly well-suited to conveying the complexity and structure of microbiome datasets within an immersive 3D environment. The work also considers the added value that VR can offer in comparison to traditional desktop-based approaches. Furthermore, the paper addresses the technical challenges that may arise during the development process, ranging from performance limitations to data integration issues.
The contribution of this paper is the design and implementation of a VR application called MicroLabVR. This application should serve as a proof of concept and should process and visually represent spatiotemporal microbiome datasets. The datasets should be created with a selected simulation tool. MicroLabVR should be designed by considering various design principles to optimize the user interface (UI) and user experience (UX) design.\\
Section \ref{sec:forschungsstand} first explains the state-of-the-art of current software for microbiome modeling, focusing on two modeling approaches in particular. The section also covers common methods of microbial data visualization. Building on this, Section \ref{sec:methodology} introduces the methodology, detailing how the microbial datasets are generated and structured. It explains the concept of the tests used to evaluate MicroLabVR's performance and discusses its data architecture. Subsequently, the implementation of the VR application is described in Section \ref{sec:implementierung}, along with the associated performance optimizations. The following section outlines the performance of the application and its impact on the dataset size. To reflect on the results, the discussion in Section \ref{sec:diskussion} examines key findings. Finally, the work concludes with an outlook on future research on the topic.

\section{State of the art}  \label{sec:forschungsstand}
Research on microbial populations in recent decades has shown the importance of microbiomes for the environment and human health \cite{Bauer2017}. Abnormal compositions are associated with various human diseases \cite{Lozupone2012}.\\
Current omics methods allow only limited insights into the underlying functions of the microbiome. In particular, individual and shared metabolic functions of the species and interactions between them are rarely determined. At this point, mathematical models have shown their potential to investigate complex systems such as microbiomes \cite{Kumar2019, Fabien2009, Tomlin2007}.

\subsection{Software for microbiome modeling}
Since MicroLabVR relies on simulation software to generate suitable datasets, it is important to explore different modeling softwares. This allows us to understand why we chose BacArena \cite{Bauer2017} to complement MicroLabVR. As aforementioned, two methods of modeling spatial and temporal microbiome development are CBM and ABM. While both strategies have their advantage and limitations, both methods were used to obtain relevant findings. Furthermore, it is of interest which visualization methods are currently used and why we need to talk about the suitability of different approaches.

\subsubsection{Constraint-based modeling software} \label{sec:cbm}
Models in the context of CBM have proven their worth in simulating metabolic networks of microbial communities. They offer the opportunity to investigate microorganism-host interactions and to understand reaction pathways \cite{Thiele2010}.\\
Genome-scale metabolic models (GEMs) form the basis for restriction-based models. They contain all biochemical reactions of an organism and serve as a mathematical representation of its metabolic network. The stoichiometric matrix, as explained in Section \ref{sec:introduction} on FBA, is the most important component of such a GEM. This mathematical representation captures the stoichiometric relationships between the metabolites and the biochemical reactions \cite{Kumar2019}.\\
While some models focus on the optimization of one species, MICOM (short for: Microbial Community) extends the approach of metabolic modeling to the community level \cite{Diener2020}. Among other things, the model takes abundances into account in order to represent the uptake of metabolites more precisely. The logical conclusion is that the more abundant a species, the greater the uptake of present substances and the release of end products. For a better prediction, realistic information about the diet and cross-feeding interactions was also included \cite{Diener2020}.\\
Diener et al. \cite{Diener2020} tested their model on data from diabetes patients and a control group (n = 186), assuming a typical Western diet. The result was that bacteria showed strong variations in growth among the groups. Inter-microbial interactions were specific to the respective communities. Competitive interactions predominated, indicating competition for limited resources. The model also showed that gut species involved in the maintenance of health had a dominant set of interactions. In addition, the expected lower production of short-chain fatty acids was observed among diabetics.\\
Although FBA enables the prediction of metabolic fluxes and growth rates within a microbial community under the assumption of a steady state, this does not allow the dynamic changes that usually occur in a microbiome to be considered. COMETS (short for: computation of microbial ecosystems in time and space) is based on the dFBA (see Section \ref{sec:introduction}), whereby the metabolic processes within the cell are in a steady state, but the abundances of the organisms and the occurrence of metabolites are dynamic. In this way, temporal changes in the population can be observed and studied \cite{Dukovski2021}.\\
One of the main functions of COMETS is the simulation of bacterial growth. This function was demonstrated by the developers recreating the well-known experiment by Varma and Palsson \cite{Varma1994} to study \textit{Escherichia coli}. COMETS also allows simulation in the chemostat, which ensures consistent conditions. Another feature is that the properties of the environment can be changed, such as taking the day-night cycle into account. This is particularly useful for studying organisms that photosynthesize \cite{Ofaim2021}. Other procedures include, e.g., the simulation of extracellular reactions and the simulation of evolutionary processes \cite{Dukovski2021}.\\
One limitation of COMETS is that it inherits the limitations of FBA, such as the lack of gene regulation. This means that it is not possible to make dynamic changes to gene expression. The fixed time steps can also cause numerical errors. To minimize the susceptibility to errors, the time steps should, therefore, be chosen small enough. A simulation typically takes 30 minutes to a few hours. However, if the parameters are chosen too small, the simulation times can be several days \cite{Dukovski2021}.

\subsubsection{Agent-based modeling software} \label{sec:abm}
\begin{figure}[h]
    \centering    \includegraphics[width=0.435\linewidth]{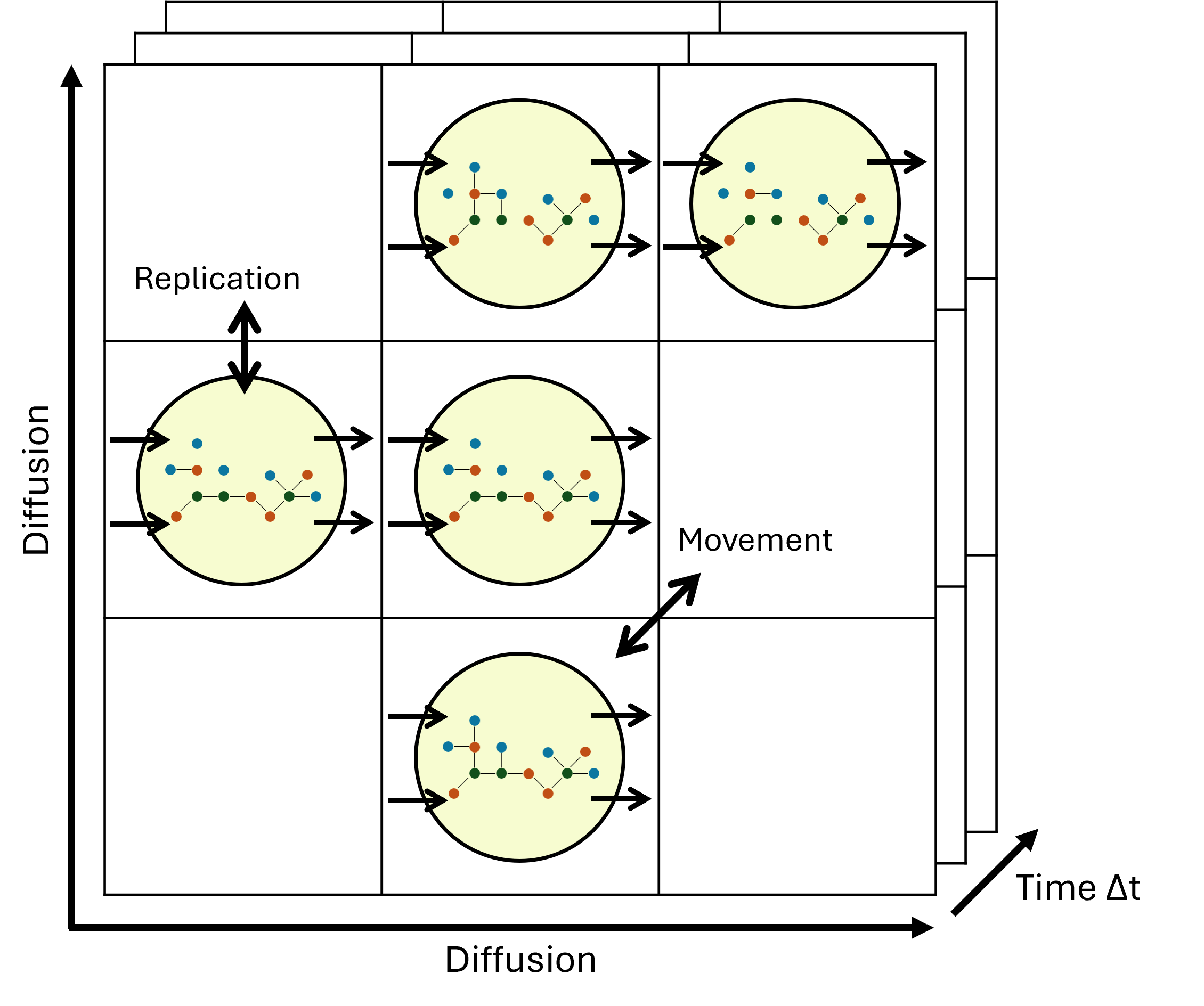}
    \caption[Simulationsgitter in BacArena]{The individual organisms are distributed on a 2D grid (modeled after \cite{Bauer2017})}
    \label{fig:grid}
\end{figure}

As has been shown so far, CBM has proven its worth in investigating and understanding the metabolic processes of microbial communities. FBA, in particular, plays a key role here. Applications of ABM relate more to complex interaction dynamics between microorganisms, as they capture the activities of single individuals \cite{Bauer2017}.\\
MatNet \cite{Biggs2013} combines the two modeling methods, CBM and ABM, allowing local interactions of microorganisms in a community to be simulated. However, one limitation of the tool is that it is restricted to monocultures \cite{Biggs2013}. Using the example of the pathogen \textit{Pseudomonas aeruginosa}, Biggs and Papin \cite{Biggs2013} demonstrated the importance of such a software extension. They reproduced the researched structure of the biofilm of \textit{P. aeruginosa}, which is created by the restricted diffusion of oxygen.\\
Bauer et al. \cite{Bauer2017} have also dedicated themselves to the task of linking both approaches and published the R package BacArena\footnote{\url{https://cran.r-project.org/src/contrib/Archive/BacArena} (last accessed 10.08.2025)}. The package extends MatNet's approach to modeling polycultural communities. The simulation area of BacArena is a 2D grid on which the individual microorganisms are distributed (see Figure \ref{fig:grid}). During the simulation, the distribution of organisms can change between the temporal sections by migrating, dying, or duplicating in a neighboring grid cell. In this way, temporal dynamics of the microbiome are represented \cite{Bauer2017}.

\begin{figure}[h]
    \centering
    \begin{subfigure}[b]{0.35\textwidth}
        \centering
        \includegraphics[width=1\linewidth]{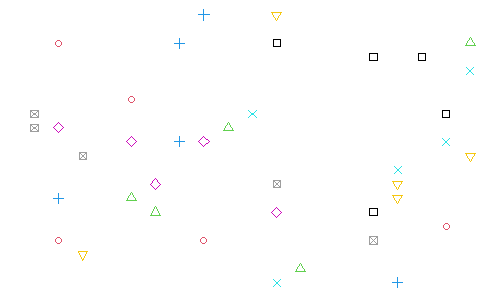}
        \caption{After one hour}
    \end{subfigure}
    \begin{subfigure}[b]{0.35\textwidth}
        \centering
        \includegraphics[width=1\linewidth]{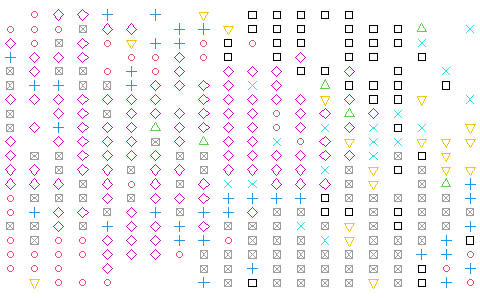}
        \caption{After seven hours}
    \end{subfigure}
    \caption[Sihumi Population Development in BacArena]{The development of eight gut bacteria growing for eight hours on a Western diet (created with BacArena \cite{Bauer2017})}
    \label{fig:sihumi_pop}
\end{figure}

\begin{figure}[h]
    \centering
    \begin{subfigure}[b]{0.35\textwidth}
        \centering
        \includegraphics[width=1\linewidth]{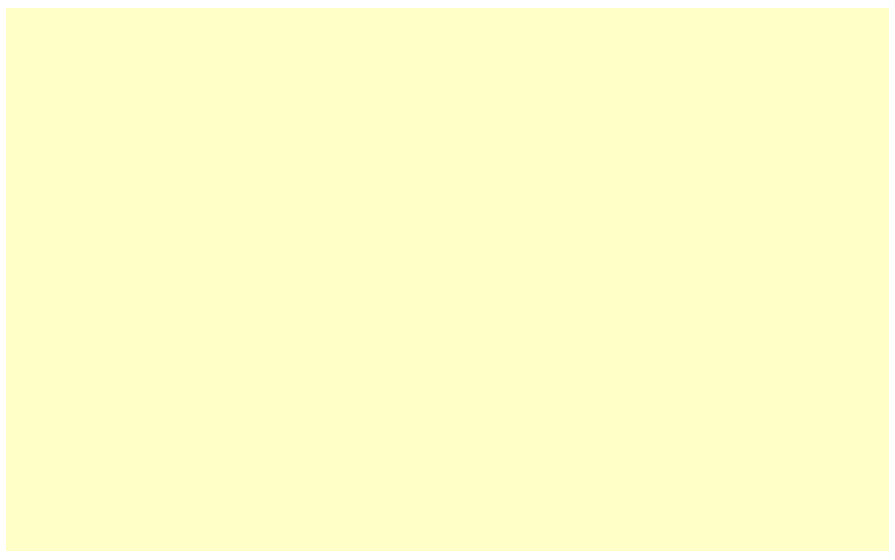}
        \caption{After one hour}
    \end{subfigure}
    \begin{subfigure}[b]{0.35\textwidth}
        \centering
        \includegraphics[width=1\linewidth]{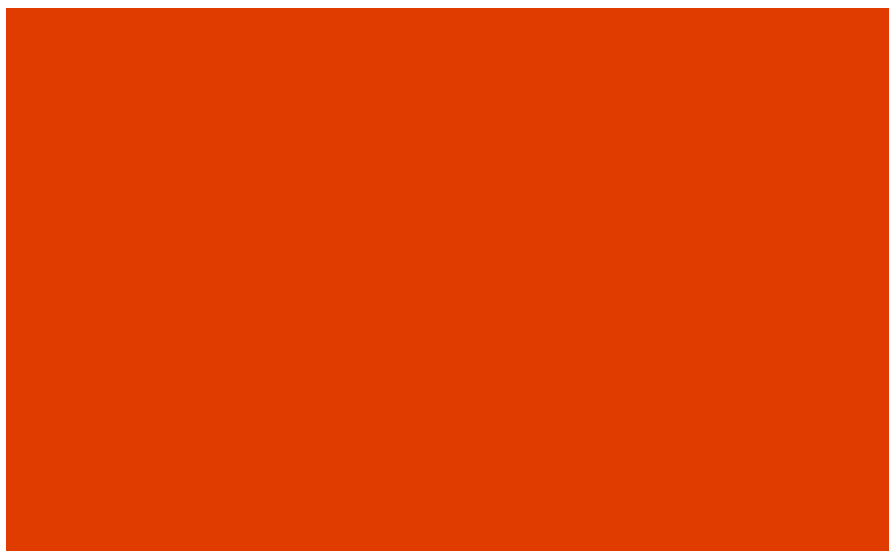}
        \caption{After seven hours}
    \end{subfigure}
    \caption[Sihumi Ammonium Development in BacArena]{The development of ammonium concentration in the same simulation area. The darker the color, the higher the concentration (created with BacArena \cite{Bauer2017})}
    \label{fig:sihumi_sub}
\end{figure}

\noindent Substance concentrations are assigned to the environment. These can be absorbed in a metabolic process, depending on the state of the organism. Organisms can also produce biomass themselves. The metabolism is calculated for each agent in the simulation using the FBA and the associated GEM, as well as the amount of biomass produced. The package allows the import of established or user-defined GEMs. The prerequisite is that the models are based on the Systems Biology Markup Language\footnote{\url{https://sbml.org/} (last accessed 14.08.2025)} \cite{Hucka2003, Hucka2018}. The biomass is calculated from the concentration of metabolites in the environment and the position of the organism in the simulation area. The metabolite concentration changes at the affected location and is calculated from the difference between the consumed and produced metabolites. This creates a separate metabolic profile for each agent, a so-called "metabolic phenotype". In this way, it is possible to access the mutual supply of metabolites between the individuals \cite{Bauer2017}.\\
All simulation steps can be visualized in order to compare them with previously collected data from experiments. Figure \ref{fig:sihumi_pop}, for example, shows the development of eight gut bacteria that have grown for eight hours on a Western diet. Each symbol represents a taxon. Correspondingly, Figure \ref{fig:sihumi_sub} illustrates the development of the concentration of ammonium.

\noindent The developers state that, compared to MatNet, BacArena allows the simulation of a community that consists of a large number of species and is, therefore, not limited to just one species. In addition, there is a function of setting kinetics for exchange reactions according to Monod \cite{Luong1987, Pinto2017}. Another feature is the consideration of phenotypes in the simulations, which takes into account the heterogeneity of the microbial communities. According to the developers, the runtime is linear and remains constant beyond a number of 50 species. The computing time is supported by parallel processing of the data. In contrast to MatNet, \mbox{BacArena} as an R package does not offer a GUI \cite{Bauer2017}. All commands and functions of \mbox{BacArena} are documented \footnote{\url{https://rdrr.io/cran/BacArena/man/} (last accessed 14.08.2025)}.\\
Bauer et al. \cite{Bauer2017} reported in the same paper on the development of BacArena on its application for modeling the biofilm formation of \textit{P. aeruginosa}. They showed the differentiated arrangement of individuals with different metabolic phenotypes resulting from the underlying nutrient distribution. The phenotypes were distributed in areas depending on the occurrence of glucose, oxygen, acetate, succinate, and carbon dioxide. The occurrence of phenotypes occurred in all simulation steps, and the population size correlated with higher nutrient concentrations. At the same time, they also showed cross-feeding between some organisms. More specifically, they found that end products of the fermentation process of some phenotypes are consumed in favor of other members, thereby promoting their growth \cite{Bauer2017}.\\
In a further simulation, they investigated the behavior of seven bacterial species, which were intended to represent a simplified human intestinal microbiota. The simulation showed an uneven distribution of mucus glycans. This had the effect of creating certain areas that were particularly suitable for bacteria that were able to break down these glycans. These bacteria found an optimal environment in which they could grow and multiply. Their investigations, therefore, indicate that gradients created by metabolic processes play a significant role in the composition of the intestinal microbiota. Bauer et al. \cite{Bauer2017} concluded from this: "This has some important implications in understanding the mucus barrier and indicates that dietary or metabolic treatments might be more relevant than immunosupressors [sic] in case of a disrupted mucosal microbiota" \cite[p. 10]{Bauer2017}.

\subsection{Microbial data visualization} \label{sec:datenvisualisierung}
In order to be able to interpret data, especially in times of big data, it is important to present it adequately. The correct visualization of data is the key to improving the viewer's perception \cite{Donalek2014}. Spatial data, in particular, which describes the structure of a microbiome, can only be displayed in 2D on a computer screen. A diagram that expands the representation space is the so-called heat map (see Figure \ref{fig:sihumi_sub}). Heat maps have the advantage that they can not only represent an area distribution, but the coloring of the data points also indicates an intensity value. The disadvantage is that relations between the intensities often cannot be interpreted precisely.\\
Moore et al. \cite{Moore2010} addressed this typical restriction of heat maps: "The advantage of this approach is that the third dimension provides additional layers of information that cannot be visualized using a traditional 2D heat map" \cite[p. 154]{Moore2010}. They, therefore, developed software that enables the analysis of multidimensional data on the human microbiome. They also integrated the option of a stereo view into their application, which allows a more intensive depth perception with 3D-capable monitors.\\
Garg et al. \cite{Garg2017} emphasized that the distinction between correlation and causality is crucial for understanding diseases and creating adequate treatments. As the authors illustrated with the example of cystic fibrosis, the microbiotic distribution in organs can affect the severity of the infectious disease. They used omics methods to analyze the microbiome and projected it onto CT images of the lungs using different software programs. Their results showed that the effect of pharmaceutical treatments and the occurrence of pathogens vary within the lungs. This also applies to the consequences of various environmental stresses. They suspect that such differences may be the cause of the severity of the disease within the lung \cite{Garg2017}. With their work, Garg et al. illustrated how data visualization can help understand interactions between organisms and the host.\\
Applications like these offer the advantage of visualization in 3D, which makes it easier to create spatial references. What these desktop applications lack, however, is immersion \cite{Zheng1998, Jennett2008}. VR simulations create the effect of immersion and allow the user to better perceive the data spaces. This can lead to a better understanding of the data and its relationships \cite{Donalek2014}. A better interpretation of spatial relationships is particularly important for complex biomolecular network data such as GEMs. This is because the datasets increase in size with further insights into biological processes, such as through high-throughput sequencing methods, and thus make their evaluation and analysis more difficult. Aichem et al. \cite{Aichem2022} therefore developed a hybrid application that offers both a desktop environment and a VR interface. In this way, they wanted to use the advantages of both tools. On the one hand, they wanted to offer a better 3D representation and, on the other hand, protect the user from cybersickness through shorter VR use. Network structures of metabolic models of an organism can be interactively visualized in the VR environment. The representation can be hierarchically manipulated and scaled so that the data can be viewed from different perspectives to better address spatial memory \cite{Aichem2022}.\\
The limitations of desktop applications were also addressed in a study by Pirch et al. \cite{Pirch2021}. They referred in particular to the complexity of the networks, which quickly look like confusing clusters on computer monitors, making analysis almost impossible. VRNetzer \cite{Pirch2021} was developed to improve the visualization and analysis of such networks, as VR technologies enable better depth perception. This means that even nodes that are obscured on a 2D plane can be visually distinguished in a 3D view. As a proof of concept, they showed how VRNetzer can be used to explore molecular networks in order to investigate links between genes and various diseases.

\section{Methodology} \label{sec:methodology}
To develop MicroLabVR as a tool for visualizing microbiome simulations, a structured methodology was essential. This includes converting simulation data into a format compatible with VR, designing a well-structured data architecture, and testing the application's performance.

\subsection{Dataset converter and use of VR}
This work uses BacArena to simulate spatial and temporal microbiome development since it combines CBM and ABM to allow the study of both microbial metabolism (CBM) and complex dynamics between interacting organisms (ABM) on a community level. BacArena allows easy access to important data structures of the executed simulation. Because it is a package for R, we designed a converter in R that transforms these relevant data volumes into CSV datasets, which can be processed by the developed VR application.\\
The converter is intended to help the user produce datasets that can be read directly into MicroLabVR. By default, it uses the demo dataset SIHUMI\footnote{\url{https://rdrr.io/cran/BacArena/man/sihumi_test.html} (last accessed 13.08.2025)}. It is a test dataset of eight gut bacteria (\textit{Anaerostipes caccae}, \textit{Bacteroides thetaiotaomicron}, \textit{Bifidobacterium longum}, \textit{Blautia producta}, \textit{Clostridium butyricum}, \textit{Clostridium ramosum}, \textit{Escherichia coli}, \textit{Lactobacillus plantarum}), which were grown over a period of eight hours on a Western diet. The SIHUMI dataset is selected as it provides a manageable set of species and number of bacteria for demonstration purposes. The substances are limited to only the six most fluctuating substances to reduce the dataset size. Population data includes time steps, biomass, genotype, and phenotype, as well as the x and y coordinates of each organism, while for substances, a concentration matrix is returned per time step. Due to limitations in direct flux mapping per organism in BacArena, randomized flux values between -50 and 50 were introduced to approximate metabolic activity in the dataset. The finalized converter generates two datasets: the “population\_dataset.csv” and the “substance\_dataset.csv”. Let us have a closer look at the structure of both datasets.\\
\textbf{\textit{Population dataset:}} The first column contains the label of the dataset: "Population". This serves as a security mechanism so that it can be checked during subsequent processing whether it is the correct dataset. The next columns contain the time to which the entry belongs and the coordinates at which x and y location the organism is located. This is followed by the biomass of the microorganism and its numerical genotype and phenotype. The eighth column contains the full name of the organism, which is assigned to the genotype. The name is made up of several fragments separated by underscores: \textit{genus\_species\_strain}. The final six columns form the flux information for the six metabolites.\\
An entry in the dataset can look like this:
\begin{figure}[h]
    \centering    \includegraphics[width=0.75\linewidth]{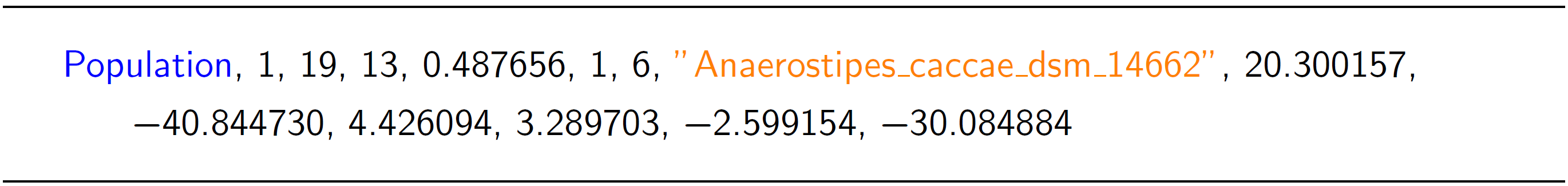}
    \caption[Population dataset structure]{Example of one row in the population dataset}
    \label{fig:pop_structure}
\end{figure}

\textbf{\textit{Substance dataset:}} Unlike the population dataset, a usable entry in the substance dataset is made up of several rows. The concentration matrices are split line by line. This means that the label "Substance" appears first. The next entry is the substance name and the timestamp. This is followed by the row number of the concentration matrix that belongs to the named substance and time. All concentration values of a matrix row follow, depending on the x-dimension of the simulation area. Figure \ref{fig:subs_structure} illustrates a 4\,x\,5 matrix:
\begin{figure}[h]
    \centering    \includegraphics[width=0.75\linewidth]{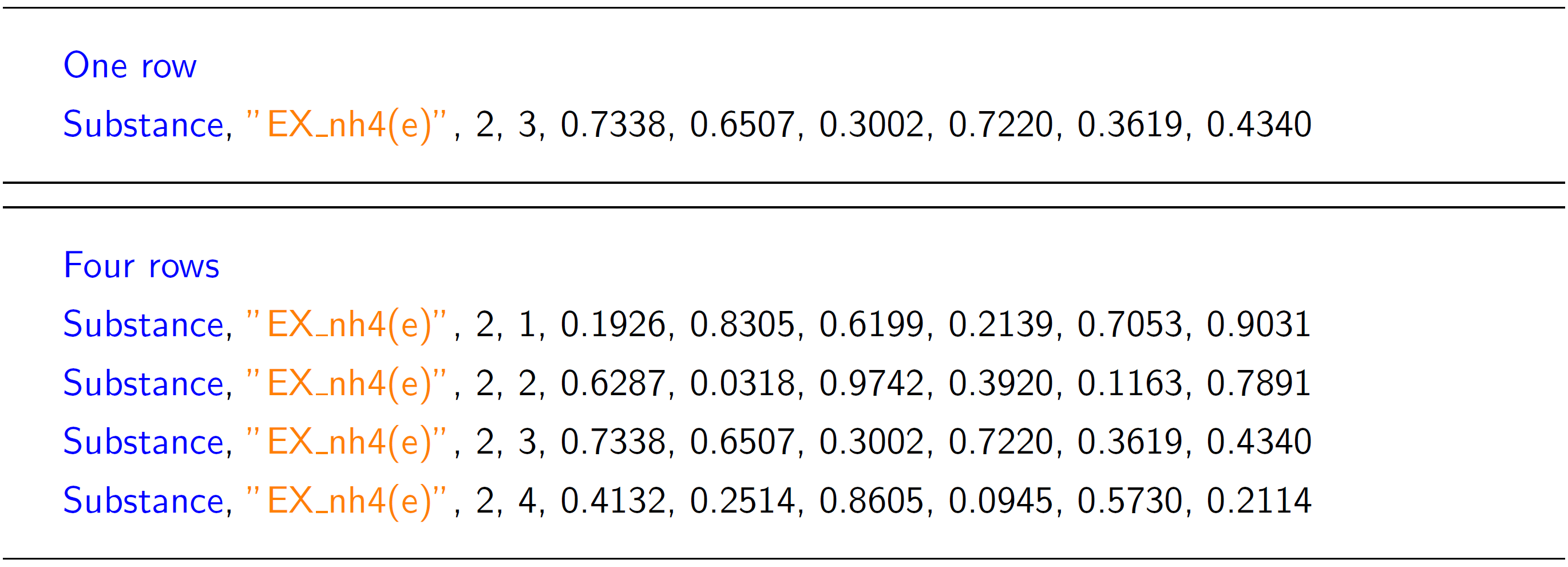}
    \caption[Substance dataset structure]{Example of one and four rows in the substances dataset}
    \label{fig:subs_structure}
\end{figure}

The entries can be read as follows: The concentration of ammonium at time 2 is 0.4132 millimolar (mM) at point (1, 1) in the simulation range and 0.3920 mM at point (4, 3).\\
As part of the essential work of this paper, we leverage the benefits of VR to improve the visualization of datasets that record microbial development. In order to test and debug MicroLabVR, the Meta Quest 3 in the 512 GB memory version from Meta, with a Snapdragon XR2 Gen 2 processor from Qualcomm \cite{Meta2024b}, is used. It is the current state-of-the-art standalone HMD with inside-out tracking, and thus, allows freedom of movement. This is because it does not need to connect to a PC via cable to enter PC VR mode, to benefit from the PC's higher computational power to run more graphically demanding applications. HMDs like the Meta Quest 3 have built-in cameras that record the environment and use this data to determine their position and orientation.\\
MicroLabVR was developed using the Unity Game Engine. The Unity Game Engine is a real-time development platform for 2D and 3D applications \cite{UnityEngine2024}. It enables the development of applications for platforms such as Windows, iOS, Linux, Android, and VR devices. The engine contains pre-implemented \textit{Render Pipelines}, which can be selected according to the developer's preferences and application requirements and affect the performance and authenticity of the graphics \cite{UnityGraphics2024}.\\
To ensure user-friendliness and usability, UI and UX design must follow established principles such as the Web Content Accessibility Guidelines (WCAG), which emphasize legibility through appropriate font sizes and color contrast \cite{Norman2013, Sherwin2015, W3C2024a, W3C2024b}; tools like the Adobe Color Contrast Analyzer\footnote{\url{https://color.adobe.com/create/color-contrast-analyzer} (last accessed 14.08.2025)} can support this process. Effective UX design integrates both visual and interactive elements, ensuring that users can intuitively control the program and understand system instructions, as outlined in ISO standards \cite{ISO_9241_110, Muslim2019, Bevan2009}.
Regarding interactive elements, Fitts' Law formulates that the time it takes a user to click on a UI element depends on the distance and width of the element \cite{Fitts1954}. Following Fitts' Law, UI elements should be strategically placed to optimize interaction efficiency, a principle that extends to VR applications \cite{Clark2020}. Reducing movement distance and increasing target size can significantly enhance usability.

\subsection{Data architecture}
At the start of software development, a well-thought-out program structure is essential in order to clearly define areas of responsibility and approach development in an orderly manner. It also makes it easier for other developers to understand class and object relations for later modifications.\\
Since users should control the VR application via UIs, a UI class must be created that has access to all UI elements, such as \textit{Buttons}, \textit{Slider}, etc. It should be able to access all other UI classes (\textit{MainUI}, \textit{SimulationUI}, \textit{TutorialUI}) that manage the functions behind the controls. The tutorial should be controlled by the class \textit{TutorialUI}. The classes \textit{MainUI} and \textit{SimulationUI} control the import of the datasets and influence the simulation process. Therefore, they must communicate with the \textit{FileManager}. The \textit{FileManager} is the heart of the application. It should take over all tasks related to data management. It is responsible for reading the datasets, checking them, and converting them into usable data formats. The \textit{FileManager} is closely related to the \textit{EntityManager}. This creates entities (organisms and substances) from the simulation data, which act as representation objects for the data. For this purpose, it should instantiate objects of the classes \textit{Microorganism} and \textit{Substance}, which contain all relevant attributes. The pseudo-UML diagram for MicroLabVR looks like this:
\begin{figure}[h]
    \centering    \includegraphics[width=.6\linewidth]{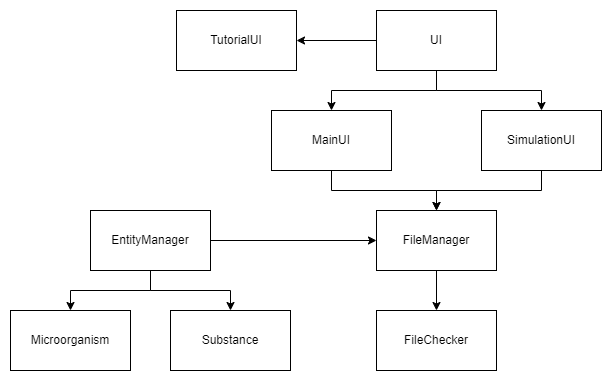}
    \caption[Pseudo-UML-Diagramm]{The data architecture of MicroLabVR. A directed arrow indicates that the class from which it originates can access the target class}
    \label{fig:uml}
\end{figure}

\subsection{Performance tests}
VR applications are generally more prone to lags than regular computer programs, as they run on weaker hardware compared to a desktop PC with a graphics processor and require twofold real-time graphical rendering. This limitation can greatly reduce the complexity and size of the datasets and, at the same time, worsen the comfort for the user by potentially leading to cybersickness. For this reason, the limits of MicroLabVR should be explored after implementation. MicroLabVR will be tested both in the standalone version and in the PC VR version in order to analyze how much the performance of the two variants differs. The PC has the following technical specifications: A 12th Gen Intel® Core™ i7-12700KF CPU, an NVIDIA GeForce RTX 4070 Ti 12 G GPU, and 32 GB DDR4 RAM.\\
Several datasets of different sizes are to be created, as various values should be measured during the tests. First, the number of rows in the population datasets and the resulting number of \textit{GameObjects}. Then we want to document how long it takes to import the datasets and set up the \textit{Scene}. Attention should be paid to how the application behaves: Does it continue to run smoothly? Do the frames stutter? Does the screen freeze? Does the application crash? However, the most important parameters are the FPS during the simulation. We strive to achieve a state where the application runs with constant 70 FPS on the Meta Quest 3, as it is recommended to minimize cybersickness \cite{Palash2025}. Based on this measurement data, which is taken via the Unity Profiler\footnote{\url{https://docs.unity3d.com/Manual/Profiler.html} (last accessed 14.08.2025)}, a recommendation for the dataset size should ultimately be made.

\section{Implementation of MicroLabVR} \label{sec:implementierung}
The VR template from Unity was chosen as the basis for the project. This already contains some configurations for VR applications, such as the XR Plug-in Management and the XR Interaction Toolkit. UI templates, which can be modified as required, are also useful.

\subsection{Data visualization}
In view of the datasets produced, it was decided to represent the organisms as simple \textit{GameObjects} in the shape of a bacterial capsule in a Petri dish. The Bacteria were modeled in Blender\footnote{\url{https://www.blender.org/} (last accessed 14.08.2025)} and are given a different colored \textit{Material} depending on the species. The substances are shown as detailed \textit{Meshes} which are generated at runtime. A \textit{Shader} colors the location depending on the concentration, creating a 2D heatmap. As mentioned in Section \ref{sec:datenvisualisierung}, heat maps are well-suited for visualizing matrices. Since VR applications include a third dimension, MicroLabVR also offers a 3D heatmap to visualize the substance concentrations. For the 3D representation, the nodes are spatially raised to create hills and valleys.
\begin{figure}[h]
    \centering
    \begin{subfigure}[b]{0.3\textwidth}
        \centering        \includegraphics[width=1\linewidth]{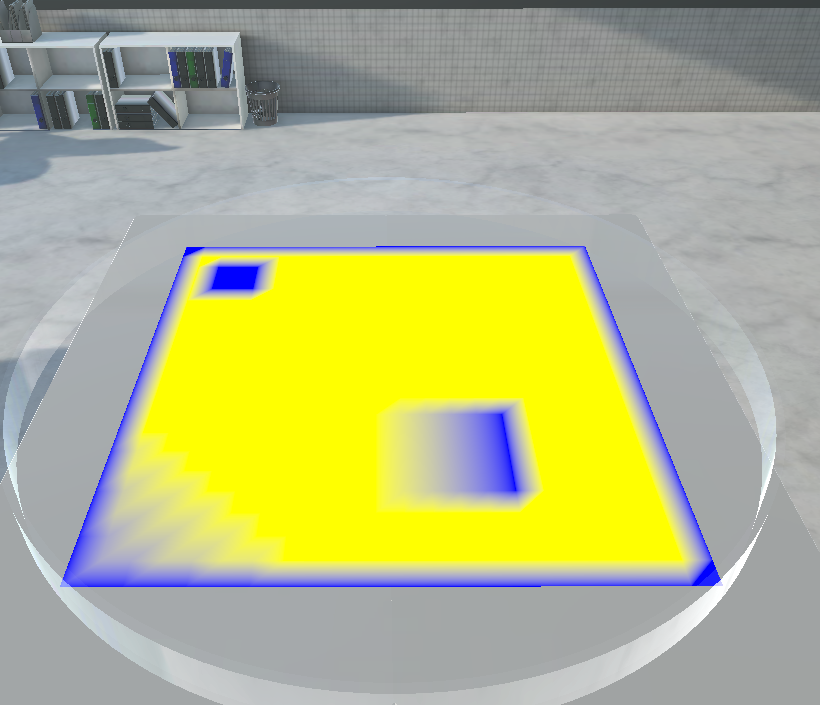}
    \end{subfigure}
    \begin{subfigure}[b]{0.3\textwidth}
        \centering        \includegraphics[width=1\linewidth]{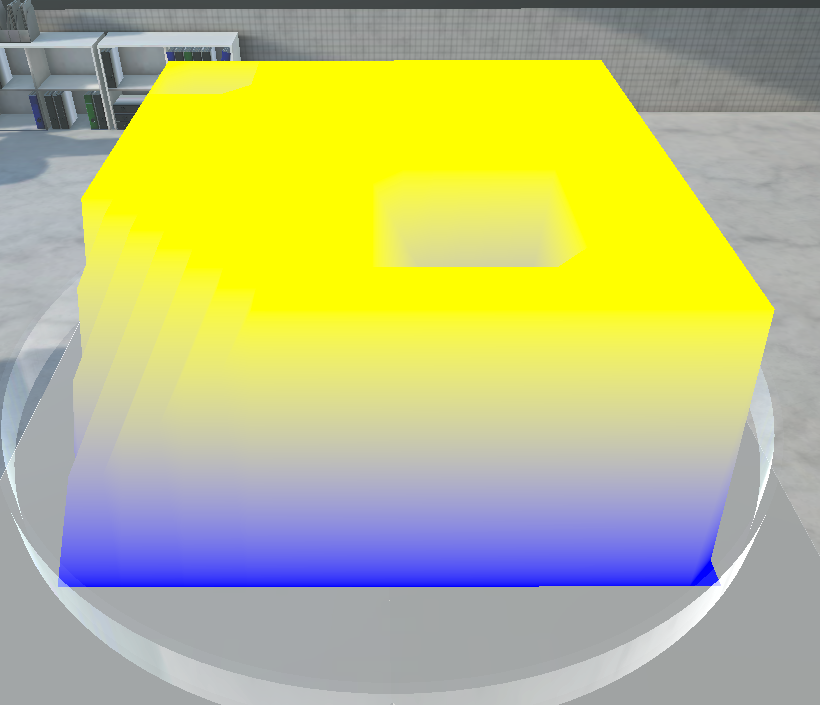}
    \end{subfigure}
\caption[Mesh of a substance]{The mesh for displaying the substance concentration in 2D (left) or 3D (right) is loaded at runtime. Blue: low concentration, yellow: high concentration}
\label{fig:substance_mesh}
\end{figure}

\begin{figure}[h]
    \centering    \includegraphics[width=0.5\linewidth]{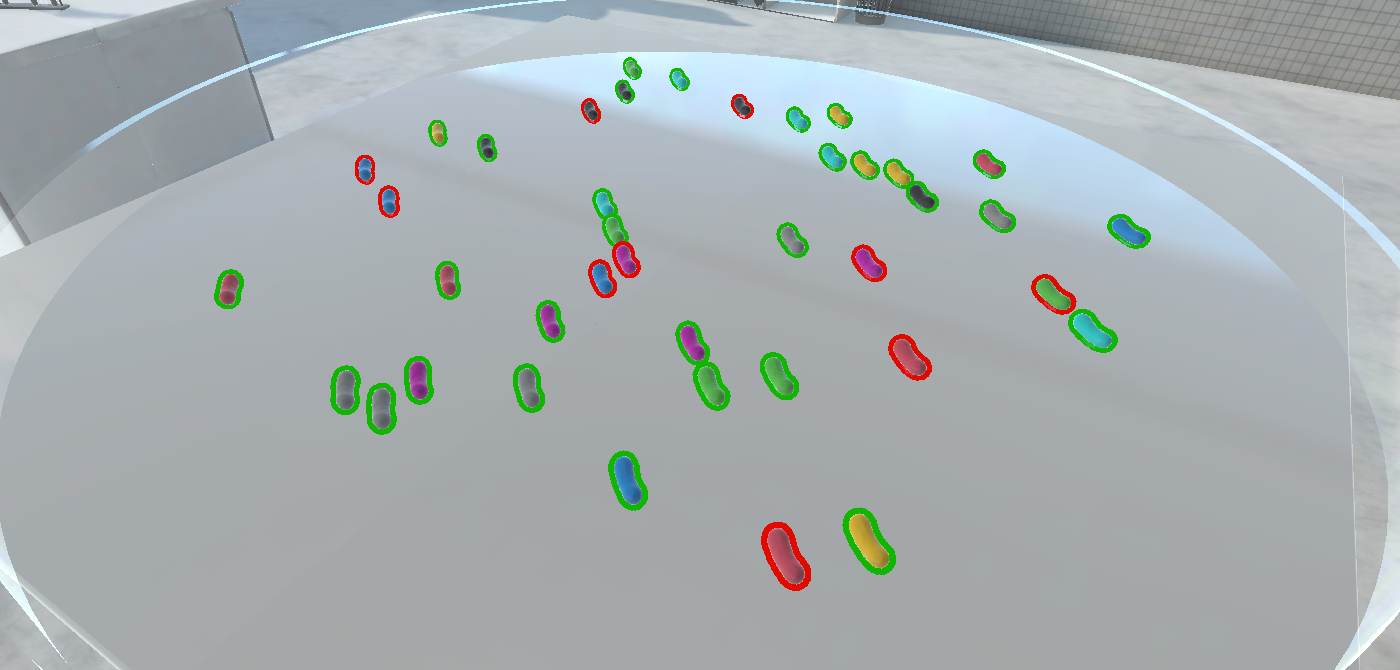}
    \caption[Flux-Outline-Visualization]{Organisms with a green border produce the substance, and those with a red border consume it}
    \label{fig:flux_outline}
\end{figure}

\noindent Whether an organism produces or consumes a metabolite is visualized using an "outline" \textit{Shader}. The 3D models are outlined in either green (production) or red (uptake) (see Figure \ref{fig:flux_outline}).

The composition of the datasets did not allow for a different method since the metabolic behavior is only represented by numerical values per organism, indicating the production and uptake rate. Missing information about which species are involved in an exchange reaction is the reason for the absence of interaction patterns.

\subsection{User tutorial} \label{sec:tutorial}
\begin{figure}[h]
    \centering
    \begin{subfigure}[b]{0.2\textwidth}
        \centering
        \includegraphics[width=1\linewidth]{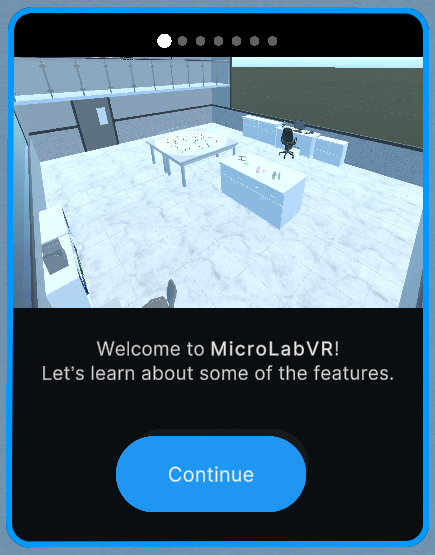}
        \caption{}
    \end{subfigure}
    \begin{subfigure}[b]{0.2\textwidth}
        \centering
        \includegraphics[width=1\linewidth]{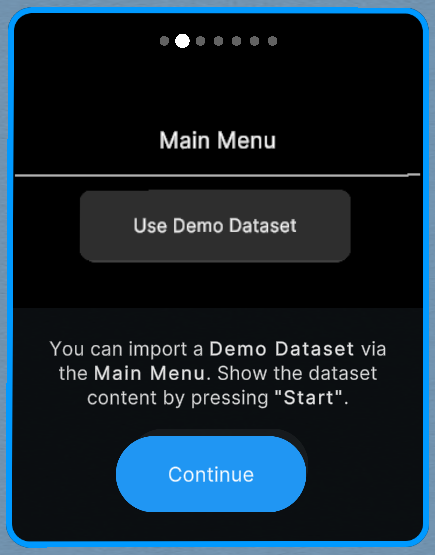}
        \caption{}
    \end{subfigure}
    \begin{subfigure}[b]{0.2\textwidth}
        \centering
        \includegraphics[width=1\linewidth]{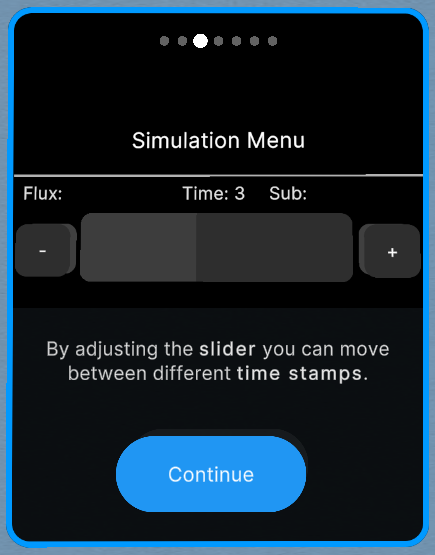}
        \caption{}
    \end{subfigure}
    \begin{subfigure}[b]{0.2\textwidth}
        \centering
        \includegraphics[width=1\linewidth]{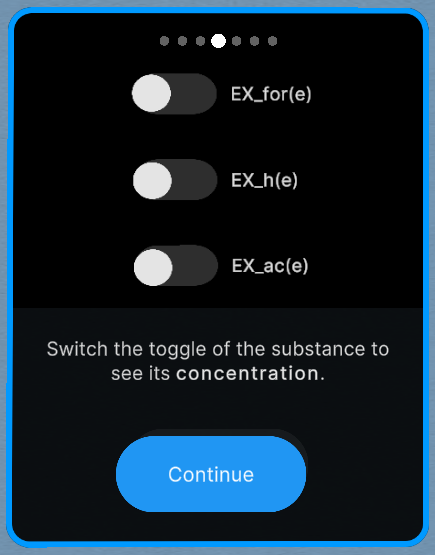}
        \caption{}
    \end{subfigure}
    \begin{subfigure}[b]{0.2\textwidth}
        \centering
        \includegraphics[width=1\linewidth]{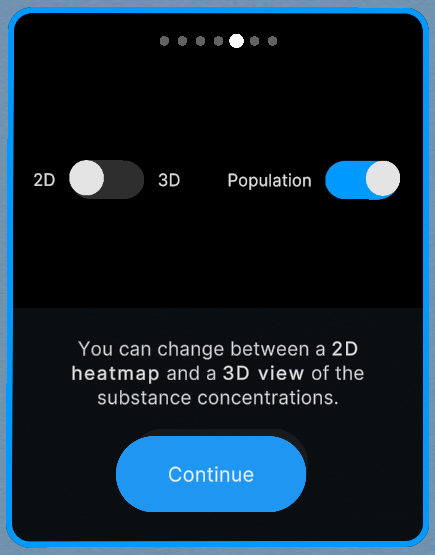}
        \caption{}
    \end{subfigure}
    \begin{subfigure}[b]{0.2\textwidth}
        \centering
        \includegraphics[width=1\linewidth]{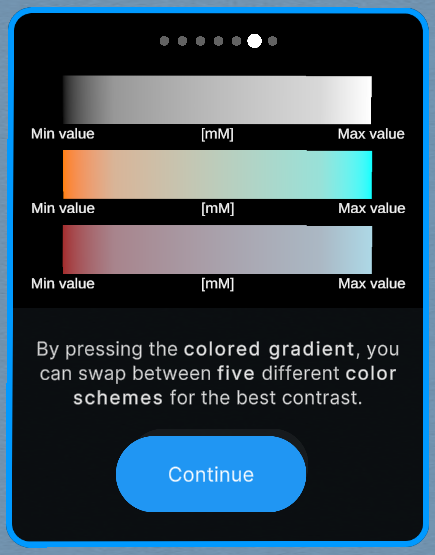}
        \caption{}
    \end{subfigure}
    \begin{subfigure}[b]{0.2\textwidth}
        \centering
        \includegraphics[width=1\linewidth]{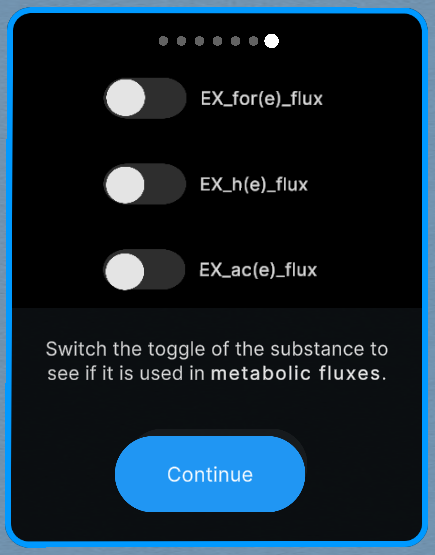}
        \caption{}
    \end{subfigure}
    \caption[User tutorial]{All slides of the user tutorial}
    \label{fig:introductionmenu}
\end{figure}

A tutorial based on a UI template of the sample \textit{Scene} was created with seven slides (see Figure \ref{fig:introductionmenu}). When entering the laboratory-like \textit{Scene}, the user can iterate through it using a \textit{Button} at the bottom of the UI. The tutorial begins with a welcome screen and a brief introduction. This is followed by instructions on importing a demo dataset, starting the simulation, and using the \textit{Slider} for time selection. Then the \textit{Toggles} for substance concentrations, switching between 2D or 3D heatmap view, and the color schemes are explained. The last slide describes the function of the flux \textit{Toggles}.

\subsection{Data management via the main menu}
\begin{figure}[h]
    \centering
    \includegraphics[width=0.3\linewidth]{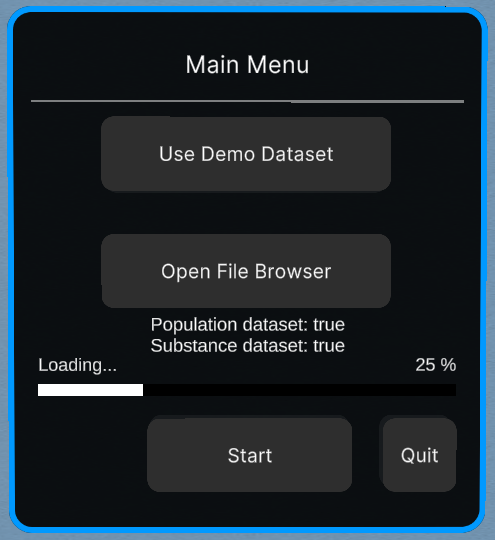}
    \caption[Hauptmenü]{The main menu of MicroLabVR}
    \label{fig:mainmenu}
\end{figure}

As soon as a user starts MicroLabVR, a main menu appears in front of them (see Figure \ref{fig:mainmenu}). The user can push the \textit{Button} labeled "Use Demo Dataset" to load the stored demo datasets from the application's resource folder. The population and substance datasets are each converted into a 2D list. The datasets can be used in this format for further backend processes. The datasets are read using \textit{Coroutines}\footnote{\url{https://docs.unity3d.com/Manual/Coroutines.html} (last accessed 14.08.2025)}. \textit{Coroutines} are methods in Unity that stretch processes over several frames. This approach allows other functions to be called while the \textit{Coroutine} is running. As a result, a progress bar could be implemented, which is updated in each frame (see Figure \ref{fig:mainmenu}).

\noindent Next, the datasets are checked to determine whether every row matches the designed rules from Section \ref{sec:methodology}. Besides valid data types, it is important to know whether the concentration matrices match the dimension of the simulation area and if the coordinates of the organisms are within this area. If there are rule violations, error messages are displayed to the user depending on the context:
\begin{itemize}
    \item \textbf{If an entry in the datasets does not have the correct format:} "Format of \textit{path} is invalid. Please check line \textit{row}, column \textit{column}. Invalid entry: \textit{entry}. Should be of type: \textit{correct format}!"
    \item \textbf{If the population dataset has more than 14 columns (as it is currently limited to six fluxes):} "Population dataset has \textit{number of columns} instead of 14 columns!"
    \item \textbf{If the duration of the simulation does not match in both datasets:} "The simulation times \textit{times} of your datasets don't match!"
    \item \textbf{If the dimensions of the simulation areas do not match in both datasets:} "The simulation dimensions of x \textit{x-dimensions} or y \textit{y-dimensions} don't match!"
    \item \textbf{If the numerical genotype cannot be clearly assigned:} "Genotype does not match a name in line \textit{row} of population dataset!"
\end{itemize}
This security measure ensures that no runtime errors occur when accessing the data if user-defined datasets are imported. The UI lists the names of the datasets paired with a Boolean value that shows the import status (true: successfully imported, false: error occurred). After the check, instances of the classes \textit{Microorganism} and \textit{Substance} are created using this data. The member variables of both classes receive the read values. Once the datasets have been successfully imported, the simulation can be loaded via the “Start” \textit{Button}. The start is also accompanied by a loading bar.

\subsection{Simulation menu}
The simulation menu is used to control the visualization of the microbiome. A \textit{Slider}, and two \textit{Buttons} labeled "-" and "+" allow the user to iterate through each timestep of the simulation. If the user does not want to look at the menu, they can simply press the assigned buttons on the left Quest 3 controller to control the timestep as well. While iterating, the Petri dish visualizes the selected state of the microbiome. During this procedure, organisms that are not currently used are pooled at a distant location. If they are needed again, these \textit{GameObjects} are not instantiated again but recycled to reduce computational effort.
\begin{figure}[h]
    \centering
    \includegraphics[width=0.3\linewidth]{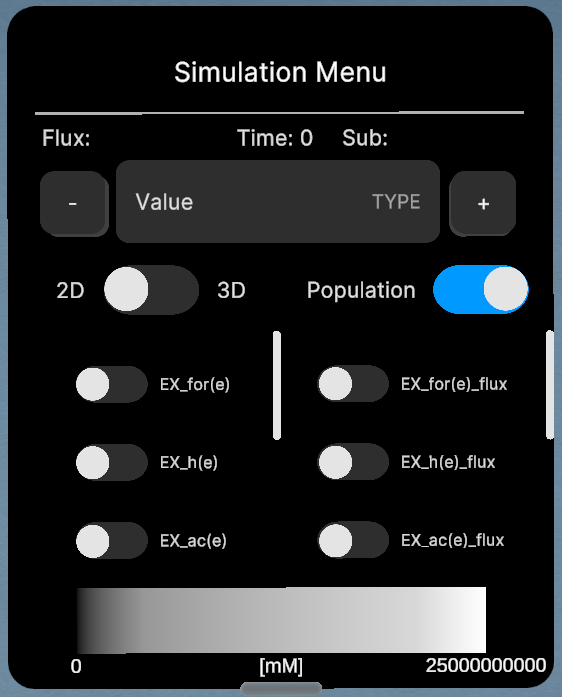}
    \caption[Simulation menu]{The simulation menu of MicroLabVR}
    \label{fig:simulationmenu}
\end{figure}

\noindent \textit{Toggles} for all available substances are listed below the \textit{Slider}. The user can select a \textit{Toggle} to display the concentration distribution. Similar to the organisms, the \textit{Mesh} of the substance is positioned in the Petri dish; all others remain invisible outside the \textit{Scene}. Only one \textit{Toggle} can be turned on at the same time, else the \textit{Meshes} would overlap. Above the list, there is another \textit{Toggle}, which is set to "2D" by default. It can be used to change the dimension in which the substance concentration is to be displayed. To achieve this effect, two \textit{Meshes} were generated for each substance after the start of the simulations - in 2D and 3D. Depending on the selected dimension, the appropriate \textit{Mesh} is displayed in the laboratory.\\
The flux \textit{Toggles} work in a similar way. However, instead of placing \textit{GameObjects} in the scene, they give the organisms the appropriate outline-\textit{Material} to indicate the production or uptake of metabolites.\\
At the bottom of the simulation menu, a color scale with a gradient shows the assigned concentration values, with the minimum value on the left and the maximum value on the right. The extreme values are extracted from the substance dataset at runtime and set in the UI. The scale itself is a \textit{Button}, which, when pressed, changes the color gradient between five color schemes.

\subsection{Interactions with organisms}
The user can interact with any organism in the \textit{Scene}. A black outline appears when the controller is moved over an organism object. A trigger press starts a hover animation for highlighting purposes. All related information, based on the attributes of the \textit{Microorganism} object, is displayed on an information panel with a microscope image of the organism. A script on each \textit{GameObject} enables the data to be accessed and forwarded.

\subsection{Aspects of UI and UX design}
As mentioned at the end of Section \ref{sec:methodology}, it is important that MicroLabVR takes aspects of good UI and UX design into account to ensure usability and user-friendliness. First of all, it was important that the user gets an understandable introduction to how MicroLabVR works so that the application meets their expectations. This is why the user tutorial was developed, which guides the user step by step through the functions of the UIs. At the same time, the tutorial contributes to self-descriptiveness.\\
When designing the UIs, particular attention was paid to ensuring that the color contrasts of the texts were intense. Contrast ratios indicate the difference in brightness between text and its background. Higher values mean better readability. The contrast ratio is calculated from the relative luminance of the text and background colors, using a formula defined by the WCAG \cite{W3C2023}. It ranges from 1:1 (no contrast) to 21:1 (maximum contrast). After adjusting the background and text color, the contrast ratios of the menus are 6.1:1 (white \#FFFFFF on grey \#626262) and 21:1 (white \#FFFFFF on black \#000000). Only the \textit{Button} of the user tutorial initially had a ratio of 3.13:1 (white \#FFFFFF on blue \#2096F3). Therefore, the color of the \textit{Button} was changed to a darker blue (\#2075B9) so that a ratio of 4.88:1 was achieved. As a result, all UIs in MicroLabVR achieve conformance level AA, as all contrast ratios are above 4.5:1 but below 7:1. If some users still have problems reading the texts or operating the buttons with the controller, they can rescale all menus larger. This function was already available in the UI templates.\\
Continuous feedback is necessary for the user if background processes are running that they have activated. In the context of the VR application, this recommendation focuses on the status display of the main menu. This progress bar shows the import status of the datasets at first and then the setup status of the \textit{Scene} and informs the user as soon as the simulation can be started. This method can prevent a user from becoming impatient or even quitting the application.\\
Visual and vibrotactile feedback can enhance the self-descriptiveness of an application. In MicroLabVR, visual feedback is used to make it clear which organism has been selected with the controller. First, it is outlined in black when hovering over it and highlighted with an animation after clicking on it. For both actions, the controller emits a brief vibration pulse. These methods can be used to avoid uncertainties as to which organism was selected. The application also emphasizes to the user which status it is in. This form of self-descriptiveness makes it clear that the objects are interactive.\\
Since MicroLabVR is to be used as a tool to analyze bioinformatic data, distractions should be avoided. It was, therefore, deliberately decided that no further interaction with the decoration in the \textit{Scene} should be possible. Likewise, the field of view of the scientist using the simulation should not be obstructed by the simulation menu. For this reason, the feature of the UIs that they can be grabbed with the controller and moved around the room has been retained. It should also be possible to hide the menu completely. Therefore, the assigned button on the right-hand controller toggles the visibility of the simulation menu. This means that the user does not have to look away from the Petri dish during the analysis process. With the same intention, the assigned buttons on the left controller were linked with the function to iterate through the time steps. It is therefore also not necessary to look at the simulation menu.\\
When it comes to UI and UX design, accessibility plays a key role. One aspect was already taken into account in the choice of color contrasts. When it came to choosing the color schemes for the concentration intensities, it was difficult to find the right color combination, as many visual impairments had to be taken into account. The final decision was to offer five schemes so that the user can choose the best contrast for them.
\begin{figure}[h]
    \centering
    \begin{subfigure}[b]{0.35\textwidth}
        \centering
        \includegraphics[width=1\linewidth]{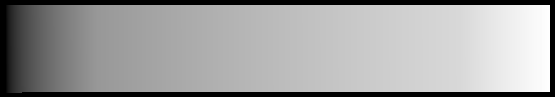}
        \caption{Black (\#000000) to white (\#FFFFFF)}
    \end{subfigure}
    \hfill
    \begin{subfigure}[b]{0.35\textwidth}
        \centering
        \includegraphics[width=1\linewidth]{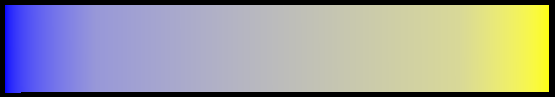}
        \caption{Blue (\#0000FF) to yellow (\#FFFF00)}
    \end{subfigure}

    \vspace{0.1cm}

    \begin{subfigure}[b]{0.35\textwidth}
        \centering
        \includegraphics[width=1\linewidth]{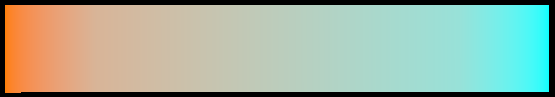}
        \caption{Orange (\#FF7F00) to cyan (\#00FFFF)}
    \end{subfigure}
    \hfill
    \begin{subfigure}[b]{0.35\textwidth}
        \centering
        \includegraphics[width=1\linewidth]{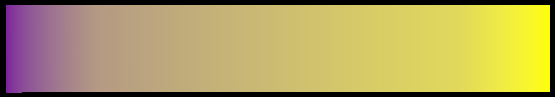}
        \caption{Violet (\#7E1E9C) to yellow (\#FFFF00)}
    \end{subfigure}

    \vspace{0.1cm}

    \begin{subfigure}[b]{0.36\textwidth}
        \centering
        \includegraphics[width=1\linewidth]{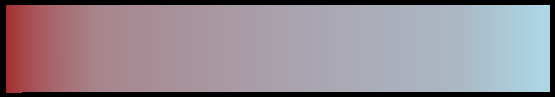}
        \caption{Brown (\#A52A2A) to light blue (\#ADD8E6)}
    \end{subfigure}
    \caption[Color schemes]{All five color schemes of the substance distribution}
    \label{fig:legend_contrast}
\end{figure}

\noindent Some propositions of Fitts' Law (see Section \ref{sec:methodology}) could also be integrated. In order to minimize the distance between the user and the simulation menu, the menu jumps into the user's field of view when the assigned button on the right controller is pressed. This feature follows the recommendation that the cursor (in this case, the center of the field of view) should be the origin of control elements. The scalability of the menus also influences the size of all \textit{Buttons} and \textit{Toggles}. Consequently, the time a user needs to navigate to them is reduced.\\
To prevent cybersickness, the option of teleportation has been integrated into the \textit{Scene}. This means that users are not dependent on moving around using a joystick. Otherwise, the passing environment could cause the associated symptoms in prone people. It was also particularly important that the standalone program runs smoothly on the Quest 3. Performance optimization methods should maximize the FPS.

\subsection{Performance optimization}
When standalone VR applications are developed, it is always a challenge to keep the computing effort at runtime as low as possible. Hence, Unity offers its own methods to increase performance. MicroLabVR uses the Universal Render-Pipeline (URP)\footnote{\url{https://unity.com/features/srp/universal-render-pipeline} (last accessed 14.08.2025)} instead of the High Definition Render-Pipeline (HDRP)\footnote{\url{https://unity.com/srp/high-definition-render-pipeline} (last accessed 14.08.2025)}. This is because high-resolution and realistic textures and effects are not necessary for a VR application in this scientific context. If the scene design required realistic \textit{Materials}, publicly available Physically Based Rendering-\textit{Materials} from the website "Free PBR Materials"\footnote{\url{https://freepbr.com/} (last accessed 14.08.2025)} were used. In this way, objects could be given more concise textures without demanding much computing power.\\
Since the user only ever sees part of the \textit{Scene}, occlusion culling\footnote{\url{https://docs.unity3d.com/Manual/OcclusionCulling.html} (last accessed 14.08.2025)} has also been integrated into MicroLabVR. Occlusion culling is a method that prevents objects from being rendered that are outside the field of view.\\
The light modes of all light sources of the \textit{Scene} have been set to "baked"\footnote{\url{https://docs.unity3d.com/2017.2/Documentation/Manual/LightMode-Baked.html} (last accessed 14.08.2025)}, which means that the lighting of the \textit{Scene} is calculated manually in advance. At runtime, however, this saves the real-time calculations of the light.\\
Other minor improvements that have been made are that the organisms do not cast shadows and that "GPU Instancing"\footnote{\url{https://docs.unity3d.com/Manual/GPUInstancing.html} (last accessed 14.08.2025)} has been activated for all organism-\textit{Materials}. The latter is a procedure to reduce the number of draw calls to the GPU. Multiple instances of a \textit{Mesh} are processed with the same \textit{Material} in one render call. This method is particularly useful for MicroLabVR, as several thousand organisms with the same \textit{Materials} are generated.

\subsection{Problems and challenges} \label{sec:problems}
During the development of MicroLabVR, some challenges arose. One of the problems was the already mentioned incorrect mapping between the organisms and the metabolic fluxes in BacArena. Therefore, the converter only creates the population dataset without inserting the fluxes. To create a complete dataset, all six columns of the flux data were filled with random values.\\
The metabolic interaction between the organisms was also not visualized. The reason for this is that it is not possible to track which organisms migrate to which location in the Petri dish or die between the time steps, as they are not marked with a unique ID. For this reason, the exchange between the organisms cannot be tracked either. This type of tracking is atypical in reality anyway because symbiotic relationships tend to exist between entire species rather than between individual organisms. Since this feature is missing, no flux gradients were visualized.\\
The aim of the application was also to enable user-defined datasets to be loaded. The functions required for this feature were also implemented and were originally intended to be executed by the “Open File Browser”-\textit{Button}. However, as access to the local file system of the Quest 3 is denied, no individual datasets can be selected. Editing the manifest file to grant access permission did not solve the problem.\\

\section{Performance tests} \label{sec:limittesting}
\begin{figure}[h]
    \centering
    \begin{subfigure}[b]{0.49\textwidth}
        \centering
        \includegraphics[width=1.0\linewidth]{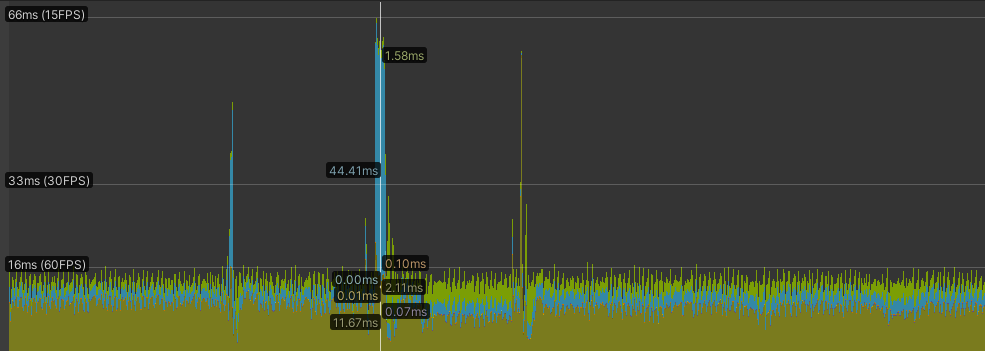}
        \caption{Meta Quest 3: $D_{Small}$}
    \end{subfigure}
    \label{fig:DatasetSQuest}
    \begin{subfigure}[b]{0.49\textwidth}
        \centering
        \includegraphics[width=1.0\linewidth]{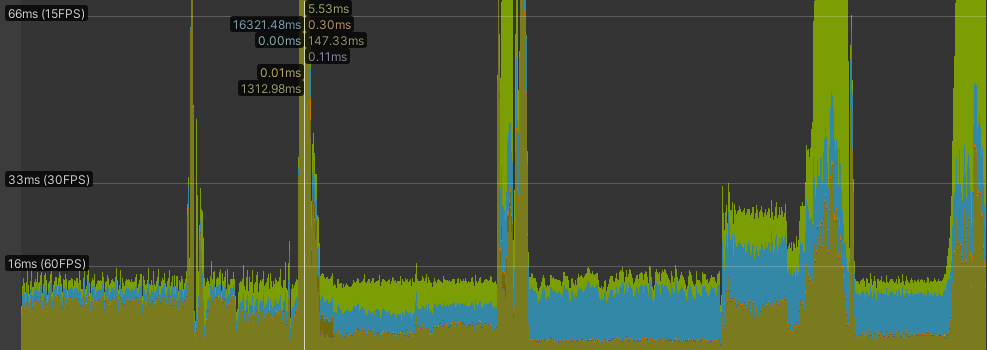}
        \caption{Meta Quest 3: $D_{Large}$}
    \end{subfigure}
    \label{fig:DatasetLQuest}
\caption[Profiler Meta Quest 3]{The history of the profiler while using MicroLabVR $D_{Small}$ vs. $D_{Large}$ in standalone mode}
\label{fig:ProfilerQuest}
\end{figure}

With regard to the potential case of importing user-defined datasets, it is interesting to know what size these datasets can have so that MicroLabVR runs efficiently. Only the population dataset is relevant in this context, as its data is used to instantiate a large number of \textit{GameObjects}.\\
Three final size variants of datasets were created, with 1389 ($D_{Small}$), 10600 ($D_{Medium}$), and 48000 ($D_{Large}$) rows (see Table \ref{tab:performance}). The performance of all datasets was monitored with the Unity profiler in a developer build of the application. Its x-axis shows each frame, and the y-axis shows the duration of the frame in milliseconds (ms) (see Figure \ref{fig:ProfilerQuest}). The same procedure was chosen for each test: First, the demo dataset was imported, the simulation was started, the maximum number of \textit{GameObjects} was generated, and then we took a look around in the \textit{Scene}. This process was first executed on the Quest 3 and then on the PC. As expected, the performance of MicroLabVR decreases with increasing dataset size on both devices (for Quest 3 performance, see Figure \ref{fig:ProfilerQuest}).

\noindent As expected, the application runs more smoothly on the PC. It was noticeable that the time to prepare the simulation (after pressing the "Start"-\textit{button}) takes sufficiently longer than the dataset import (see Table \ref{tab:performance}).
\begin{table}
 \caption{The performance of the application depending on the size of the population dataset: Dimension of the simulation area ($Dim$), number of instantiated \textit{GameObjects} ($n$), time to import the datasets via the Quest 3 ($t^Q_1$) and via the computer ($t^P_1$), time to instantiate the \textit{GameObjects} via the Quest 3 ($t^Q_2$) and the computer ($t^P_2$). FPS while the simulation is running on the Quest 3 ($FPS^Q$) and the PC ($FPS^P$). The last two columns indicate the suitability for the Quest 3 ($Q$) or PC-VR ($P$) regarding user experience.}
  \centering
  \begin{tabular}{l l l | l l l l l l | l l}
    \toprule
    $Dim$ & $Rows$ & $n$ & $t^Q_1$ & $t^P_1$ & $t^Q_2$ & $t^P_2$ & $FPS^Q$ & $FPS^P$ & $Q$ & $P$ \\
    \midrule
    20 × 20 & 1389 & 392 & 0,12 s & 0,06 s & 0,59 s & 0,3 s & 71,43 & 76,92 & \cmark & \cmark  \\
    50 × 50 & 10600 & 2500 & 0,2 s & 0,15 s & 1,3 s & 0,45 s & 33,33 & 76,92 & \xmark & \cmark \\
    100 × 100 & 48000 & 10000 & 2,2 s & 0,9 s & 162,7 s & 72,9 s & 11,21 & 71,42 & \xmark & \xmark \\
    \bottomrule
  \end{tabular}
  \label{tab:performance}
\end{table}

\noindent The FPS during the computationally intensive intervals could be calculated from the measured duration of the frames using the following equation:
\begin{equation}
 FPS = \frac{1 \text{ } second}{average \text{ } duration \text{ } of \text{ } the \text{ } frames}
\end{equation}

\noindent Several crucial aspects must be considered when deciding on a limit value for the dataset size: the loading time of the datasets and the simulation, as well as the level of FPS. During the performance tests, it was significant that $D_{Large}$ had too long a loading time for the simulation. During this time, the user had a frozen screen, which is intolerable. Although the FPS in PC VR mode is over 70, in standalone mode, it is just 11.21. Only $D_{Small}$ performed satisfactorily throughout, as $D_{Medium}$ was only able to achieve 33.33 FPS in the standalone application. This may be sufficient for some users, but should not be the standard for VR applications.\\
Finally, all datasets can be used in PC VR mode if a smooth loading screen is displayed during the import and setup of the simulation. As a standalone application, only $D_{Small}$ fulfills all performance requirements. During both loading phases, the user should be presented with a loading screen to prevent FPS drops and potential freezes from going unnoticed.

\section{Discussion} \label{sec:diskussion}
MicroLabVR includes all the designed features. All organisms are displayed in their correct constellation as interactable \textit{GameObjects}. The substance distributions and fluxes can be displayed via a UI. The metabolic flux display, in particular, has the potential for improvement. It is only shown with an outline around the organisms. A specific visualization of exchange reactions would be more informative. However, this would require more data from BacArena and a complex algorithm that calculates the exchange between entire species.\\
The application offers decisive advantages compared to a desktop application such as BacArena. First of all, \mbox{MicroLabVR} allows simple and intuitive navigation and control of the functions. A user tutorial supports the program's self-descriptiveness and learnability. Using the \textit{Slider}, the user can iterate through all time points more easily and does not have to rely on graphical plots. Potentially, the development and even the movement of some organisms can be tracked, even if this cannot be confirmed with certainty. The disadvantage is that not all time points can be displayed simultaneously. MicroLabVR is, therefore, suitable as an additional tool that displays microbial data from a different perspective.\\
A key advantage of VR is that it enables information to be displayed in the third dimension. Projecting 3D data onto 2D surfaces can limit interpretability. This advantage does not necessarily apply to MicroLabVR at the moment, as only one layer of a microbiome can be displayed. The reason for this lies in the choice of the simulation tool. Implementing the framework for importing and displaying 3D data in MicroLabVR is feasible in the future. However, it remains to be seen whether there is a need for it.\\
The 3D heat map display seems particularly useful in the VR application, as it makes direct use of a spatial environment. The user gets a better representation of the intensity values by better displaying the size ratio of the values. In addition, the population can be superimposed on the 2D heat map, which in some cases may reveal a migration towards a food source.\\
In addition to a better understanding of spatial relationships, users benefit from immersion by creating a realistic learning experience. The so-called "Situated Learning"\footnote{More about Situated Learning: \url{https://ijiet.org/papers/48-R017.pdf} (last accessed 10.08.2025)} can increase user engagement and the effectiveness of learning. Hence, it is worth considering whether MicroLabVR also has the potential to teach students. Whether other academics also perceive these theoretical advantages remains to be tested.\\
UX design regulations were partially implemented. Constant attention was paid to creating an accessible application. The minimum color contrast ratio is 4.88:1. In order to achieve a level AAA, the color of the user tutorial button could be made darker. Also useful for the user are smaller features, such as the outline of an organism paired with vibrotactile feedback when it is pointed at with the controller and the hover animation when it is clicked. These forms of feedback could be further enhanced by playing sounds that match the action. Only the control elements of the UIs have already implemented this feature. A further improvement would be to extend the user tutorial to an entire guided tour through the \textit{Scene}. However, since MicroLabVR only consists of one \textit{Scene}, this seems less useful. Alternatively, an audio guide would be an improvement.\\
A shortcoming of the UX in MicroLabVR could be seen in the color selection of the bacteria and the display of the fluxes. It would be a problem for people who suffer from anomalous trichromacy (red-green color blindness), as red and green \textit{Materials} were used. For them, it would be almost impossible to distinguish whether an organism consumes or produces a substance. A color selection could only be implemented to represent the concentration of the substance.\\
As the dataset size increases, both the performance of the application and the size of the organisms decrease. Even with a simulation area comprising 100\,×\,100 bacteria, it can become difficult to click on the \textit{GameObjects} according to Fitts' Law. Therefore, a lower limit for the organism size would have to be set to ensure controllability. This change would also mean that the Petri dish and thus the entire \textit{Scene} would have to become scalable.\\
Various technical problems arose during the development of MicroLabVR. Initially, it was planned that MicroLabVR would also be able to read user-defined datasets in order to extend the application's usefulness. As already described in Section \ref{sec:problems}, the Quest 3 does not allow access to the file system, which is why the datasets stored there could not be imported. An alternative solution could be to use the Storage Access Framework\footnote{\url{https://developer.android.com/guide/topics/providers/document-provider} (last accessed 15.08.2025)} for Android developers. In this way, specific shared directories ("Downloads" or "Documents") could be made available. A comprehensive solution would be to connect a cloud storage system where all datasets are stored. The challenge would be to develop a suitable interface between the application and the storage. Apart from this, file access via the PC would be easier. There are several \textit{Assets} that can be considered to integrate a file browser. The standalone file browser\footnote{\url{https://github.com/gkngkc/UnityStandaloneFileBrowser/tree/master?tab=readme-ov-file} (last accessed 30.01.2025)} was also integrated into MicroLabVR at the beginning, as it was to become a standalone application, but was discarded due to technical problems. The framework for this still exists.\\
Another problem was that the mapping of the flux data from BacArena is incorrect. The developed converter, therefore, only generates datasets that do not show metabolic fluxes. For this reason, the VR application is limited to only six substances and fluxes. If files with more substances were to be read, the data format checker would intercept them.

\section{Future work} \label{sec:fazit}
With regard to functional extensions of MicroLabVR, an expansion to 3D layers would be of interest. This would allow the full potential of VR to be exploited. The only problem would be the visualization of the substances. This is because it would not only be necessary to depict a concentration of a surface but also a volume. It would also be exciting if the microbiomes could be mapped to a place of origin, such as the lungs or the intestine. This would make it possible to identify regional differences in larger datasets.\\
One visualization method that falls somewhat short is the representation of the flux distribution. The organisms are merely color-coded to indicate the uptake or production of a substance. It would be more informative to illustrate flux directions and intensities. In this way, entire gradients could be displayed, which could improve microbiome data analysis.\\
An initial performance test showed that the datasets can initially only be very small (less than a total of 10,000 instantiated organisms). MicroLabVR already follows an approach for asynchronous loading of the datasets via \textit{Coroutines}. Nevertheless, there are FPS drops during the loading time. This shortcoming suggests that the \textit{Coroutine} logic is not yet fully developed and needs to be optimized. An alternative self-implemented solution for asynchronous loading (using pre-built \textit{Assets}), paired with proper data streaming from BacArena itself, would boost performance even more. Based on these advances, performance and UX could be improved. Although a loading bar is integrated to give the user feedback on the application's status, an additional option would be to completely isolate the user from the action with a loading screen.\\
This basic idea would improve the loading performance but not the frame rate during the simulation. One improvement approach is to optimize the designed 3D model of the organisms. It currently has 832 polygons. With 10,000 \textit{GameObjects}, that would equal 8,320,000 polygons that need to be rendered. The number could already be reduced with simple optimization methods in Blender.\\
In addition to fixing the technical problems described in Section \ref{sec:problems}, one suggestion would be to improve usability by paying closer attention to the color selection of the organisms. Especially in the future, if larger datasets are to be imported, the color assignment must follow an algorithm that generates the best possible color combination, given the number of species. Different shaders and textures could further simplify the differentiation.\\
As soon as BacArena has fixed the mapping of the fluxes to the organisms, the R converter can also be expanded appropriately. Currently, the user must also independently set the storage location and the model in the R converter. For ideal use, however, a simple, usable interface should be created that enables direct data exchange with MicroLabVR and does not require any programming knowledge. This would reward users of the VR application with an even better UX.\\
MicroLabVR serves as a proof of concept and confirms the feasibility and basic functionality of the extended visualization of microbial data. It remains to be seen whether other scientists support the discussed advantages of the application. Controlled expert interviews would therefore be useful in order to further explore the benefits of such an application in practice and possibly in teaching.

\bibliographystyle{unsrt}  
\bibliography{main}

\end{document}